\RequirePackage[table]{xcolor}
\documentclass{PoS}
\usepackage{graphicx}
\usepackage{amssymb}
\usepackage{fontenc}
\usepackage{times}
\usepackage{mathptmx}
\usepackage{amsmath}

\definecolor{darkgreen}{rgb}{0.0, 0.5, 0.0}

\makeatletter
\newcommand{\doublewidetilde}[1]{{%
  \mathpalette\double@widetilde{#1}%
}}
\newcommand{\double@widetilde}[2]{%
  \sbox\z@{$\m@th#1\widetilde{#2}$}%
  \ht\z@=.9\ht\z@
  \widetilde{\box\z@}%
}
\makeatother

\newsavebox{\mysaveboxM}
\newsavebox{\mysaveboxT}

\newcommand{\dd}{\mathrm{d}}

\newcommand{\be}{\begin{equation}}
\newcommand{\ee}{\end{equation}}
\newcommand{\sfrac}[2]{{\textstyle\frac{#1}{#2}}}

\def \bea{\begin{eqnarray}} 
\def\eea{\end{eqnarray}}
\def\bse{\begin{subequations}}	
\def\ese{\end{subequations}}
\def\bal{\begin{align}} 
\def\eal{\end{align}}

\def\mc{\mathcal}

\def\bi{\begin{itemize}} 
\def\ei{\end{itemize}}

\makeatother

    \def\d{\delta} 
\def\e{\epsilon} 
  \def\h{\eta} \def\k{\kappa}
\def\l{\lambda}  \def\m{\mu}
\def\n{\nu} \def\o{\omega}   
\def\s{\sigma}   
  
\def\O{\Omega}

\def\one{\mbox{1 \kern-.59em {\rm l}}}

\title{\begin{flushright}\small{\textnormal{RBI-ThPhys-2020-11}}\end{flushright} \vspace{25pt}
	Graded Geometry and Tensor Gauge Theories}

\ShortTitle{Graded Geometry and Tensor Gauge Theories}

\author{Athanasios Chatzistavrakidis\\
        Division of Theoretical Physics, Rudjer Bo\v skovi\'c Institute \\ Bijeni\v cka 54, 10000 Zagreb, Croatia\\
        E-mail: \email{Athanasios.Chatzistavrakidis@irb.hr}}
        
\author{\speaker{Georgios Karagiannis}\\
        Division of Theoretical Physics, Rudjer Bo\v skovi\'c Institute \\ Bijeni\v cka 54, 10000 Zagreb, Croatia\\
        E-mail: \email{Georgios.Karagiannis@irb.hr}}

\author{Peter Schupp\\
        Department of Physics and Earth Sciences, Jacobs University Bremen\\
28759 Bremen, Germany\\
        E-mail: \email{p.schupp@jacobs-university.de}}

\abstract{We review the construction of Lagrangians for higher spin fields of mixed symmetry in the framework of graded geometry. The main advantage of the graded formalism in this context is that it provides universal expressions, in the sense that a given Lagrangian describes the dynamics of any type of bosonic tensor field even though the corresponding explicit expressions in terms of local field components and their derivatives look rather different. Aside from free fields and their kinetic terms, we also consider higher derivative interaction terms that lead to second order field equations. For scalars, differential forms and bipartite tensors, these are identified with Galileon theories, written in a simple yet elegant form as a generalised kinetic term, and are gauge invariant by construction. For fields of spin higher than 2, we illustrate the candidate Galileon-like interactions and argue that full gauge invariance and locality cannot be simultaneously maintained. }

\FullConference{Corfu Summer Institute 2019 "School and Workshops on Elementary Particle Physics and Gravity" (CORFU2019)\\
		31 August - 25 September 2019\\
		Corfu, Greece}

\begin{document}

\section{Introduction}

Physical theories that attempt to describe natural phenomena contain a diversity of fields representing their fundamental degrees of freedom. Typically, these fields can be categorised with respect to their spin. In this work we focus on bosonic fields: The spin-0 case corresponds to scalar fields, which play a fundamental role in both particle physics and cosmology. Spin-1 fields are usually differential 1-forms and they are the key players in gauge theories. Furthermore, spin-2 fields appear in gravity, where the mediator of the gravitational interaction is indeed such a field. However, apart from the above cherry picked cases, in general there exists a lot of motivation to consider further types of fields. For instance, higher degree differential forms and their corresponding higher gauge theories appear in certain frameworks, notably in string theory. Spin-2 fields different than the graviton may appear through duality in higher than four dimensions, for example the Curtright field of mixed symmetry $(2,1)$ is dual to the graviton in five dimensions \cite{Curtright:1980yk,Boulanger2003}.  Ultimately, spins larger than 2 can also be of physical interest and they appear in the spectrum of string theory. On the other hand, fermionic fields correspond to half-integer spins, however we do not discuss them further in this work. 

The main purpose of this paper is to highlight the idea that graded geometry, even at its simplest form, appears very useful in constructing a certain class of interacting Lagrangians for any type of field. 
It is based on Refs. \cite{CKRS,CKS}, where more details and references may be found.
Graded geometry is based on the introduction of coordinates with different degrees, satisfying graded commutation relations. For example, bosonic coordinates are of degree-0 and commute, while degree-1 fermionic coordinates anticommute, a situation familiar from supersymmetry. In the present context we will restrict to degree-0 and multiple sets of degree-1 coordinates. An utility of this formalism is that any type of field as described above can be represented as a function of certain degree on a graded manifold. 

The usefulness of graded geometry is demonstrated in the construction of universal Lagrangians that capture theories for any type of field, although their local coordinate counterparts may look rather different \cite{CKRS,CKS}. In this work we focus on single field, linear gauge theories in flat space, with higher derivative interactions leading to second order field equations so as to avoid Ostrogradski ghosts. With the aid of a generalised Hodge star operator that we define and the Berezin integral, we first present a universal expression for the kinetic term for any mixed symmetry tensor field. Subsequently, we show how a generalised kinetic term can capture a certain class of higher derivative interactions for mixed-symmetry tensors with two sets of antisymmetrized indices (bipartite tensors) with the desired properties, which are identified as the analogues of scalar Galileons in the case of higher spins. Galileons were first constructed for scalar fields in \cite{NRT,DDE} and they were extended to differential forms in \cite{DDE2} and, finally, to bipartite tensors in \cite{CKRS}. This latter case generalizes Lovelock theories \cite{Lovelock}, which are identified as Galileons for the metric $2$-tensor.
In the present formalism, all shared features of this class of theories are highlighted in very suggestive expressions, even though their explicit coordinate form is typically extremely complicated. 

This paper includes two results that do not appear in \cite{CKRS,CKS}. The first is the already mentioned generalised kinetic term that unifies the kinetic and higher-derivative interacting sector of the bipartite tensor Lagrangians, see Eq. \eqref{generalised kinetic}. Note that for a special class of bipartite tensors, when their partial degrees are equal, there are additional Galileon interactions that are not included in the generalised kinetic term. We describe these ones in the graded formalism separately. The second new result refers to spins greater than 2. In the formalism employed here, it is rather simple to construct candidate Lagrangians that would correspond to Galileons in this general case, although this appears to us as a formidable task otherwise. Note, however, that it turns out to be impossible to rescue both properties of full gauge invariance and local second order field equations. Although this sounds reasonable in view of the difficulty in constructing interacting theories for higher spins, it is interesting to see that certain types of higher derivative terms pass a number of tests. Specifically, we present two types of non-trivial (meaning that they are not total derivatives) terms. The first is manifestly gauge invariant, however it leads to non-local field equations, while the second is manifestly local, however it does not respect the full set of expected gauge symmetries, but only a subset of those. Finally, we mention some further results and our conclusions in section \ref{sec4}.   

\section{Multipartite tensors in graded geometry}
Multipartite tensors generalize differential forms, in that their components contain more than one set of
antisymmetrized indices when expanded in some local coordinate system. We denote the vector space of $N$-partite tensors of degree $(p_1,\dots,p_N)$ on $M$ by $\O^{p_1,\dots,p_N}(M)$. When expanded locally, their components have a total number of $P:=\sum_{A=1}^N p_A$ indices divided into $N$ antisymmetrized sets according to their degree. 

Graded geometry is a natural framework to describe multipartite tensors. Let us consider the graded supermanifold
$$\mc M=\oplus^NT[1]M\,,$$
where $T[1]M$ is the graded tangent bundle over a manifold $M$, obtained by degree-shifting the fiber coordinates by one. For the purposes of this paper, we focus on $M$ being the $d$-dimensional Minkowski space. The graded manifold $\mc M$ is equipped with the degree-0 coordinates $x^\mu$ of the base $M$, as well as with a set of degree-1 fiber coordinates $\theta^\mu_{A}$, where $A=1,\dots,N$, one for each graded tangent bundle $T[1]M$. The degree-0 coordinates are naturally bosonic, while the degree-1 coordinates are fermionic and they satisfy  
\be \label{anticommuting}
\theta_A^\mu \theta_B^\nu=(-1)^{\epsilon(A,B)}\theta_B^\nu \theta_A^\mu\,,
\ee
for some parity $\epsilon(A,B)$. There are two interesting choices of convention for this parity. The one we adopt in the present paper---and also in \cite{CKRS,CKS}---is $\epsilon(A,A)=1$ and $\epsilon(A,B)=0$ for $A\ne B$. This means that degree-1 coordinates across different sets are commuting. (The other convention, which is more standard in the literature, reads as $\epsilon(A,B)=1$, in which case all coordinates are anticommuting. In any case, our results are not affected by this conventional choice.)
The relation between multipartite tensors and graded geometry can be established due to an isomorphism between smooth functions on $\mc M$ and $N$-partite tensors on $M$, namely
\be 
C^\infty(\mc M)|_{p_1,\dots,p_N}\simeq \O^{p_1,\dots,p_N}(M)\,,\quad C^\infty(\mc M)\simeq \bigoplus_{(p_1,\dots,p_N)\in\mathbb{Z}_+^N}\O^{p_1,\dots,p_N}(M)\,.
\ee
Note that this holds for total degree $P\leq d$. Due to the above isomorphism, an arbitrary $N$-partite tensor of degree $(p_1,\dots,p_N)$ can be expanded as
\be \label{local expansion}
\o_{p_1,\dots,p_N}(x,\theta)=\frac{1}{\prod_{A=1}^Np_A!}\o_{[\mu^1_1\dots\mu^1_{p_1}]\dots[\mu^N_1\dots\mu^N_{p_N}]}(x)\,\theta_1^{1}\dots\theta_1^{p_1}\dots\theta_N^{1}\dots\theta_N^{p_N}\,.
\ee 
The square brackets denote antisymmetrization with weight 1, which is just a result of the relations \eqref{anticommuting} satisfied by the degree-1 coordinates. Moreover, composition of bipartite tensors simply becomes pointwise multiplication of functions. For a $N$-partite tensor $\o\in\O^{p_1,\dots,p_N}$ and a $N'$-partite tensor $\omega'\in\O^{q_1,\dots,q_{N'}}$ with $N\leq N'$, the $(N+N')$-partite tensor $\o\omega' $ lives in $\O^{p_1+q_1,\dots,p_N+q_N,q_{N+1},\dots,q_{N'}}$ and can be expanded locally as the pointwise product of the local expansions \eqref{local expansion} of $\o$ and $\omega'$.

\subsection{Defining the calculus}
It is now useful to define some basic calculus on the graded supermanifold. This calculus was originally developed in \cite{BB1,dMH1,BB2,dMH2} and later translated into a graded geometric language in \cite{CKS}. To this end we consider derivatives with respect to the degree-$1$ coordinates,
\be 
\bar\theta^A_\mu:=\frac{\partial}{\partial\theta_A^\mu}\,,
\ee
which are naturally assigned degree $-1$. Furthermore, they obey exactly the same relations \eqref{anticommuting} as their unbarred counterparts, while the canonical (anti)commutation relations between them are
\be \label{CCM}
\{\bar\theta_\m^A,\theta_A^\nu\}=\d_\m^\n\qquad \text{and} \qquad [\bar\theta_\m^A,\theta_{B}^\n]=0\, \quad  \text{for} \quad B\ne A\,,
\ee
in the sign convention we have adopted (see \cite{Bonezzi:2018box} for the analogous formulas in the opposite convention and \cite{Bruce,Aizawa} for further recent work in our convention).
Using these operators, $N$ number operators $\hat p_A$ can be defined as 
\be \label{number}
\hat{p}_A:=\theta_A^\mu\bar\theta^A_\m\,,
\ee
where no summation in $A$ is implied.
Any $N$-partite tensor is an eigenvector of this operator, with the corresponding eigenvalue being $p_A$. 
In addition, the Minkowski metric and the corresponding trace operator can be cast into this formalism as the maps $\h_{AB}:\O^{p_1,\dots,p_A,\dots,p_B,\dots,p_N}\to\O^{p_1,\dots,p_A+1,\dots,p_B+1,\dots,p_N}$ and
$tr_{AB}:\O^{p_1,\dots,p_A,\dots,p_B,\dots,p_N}\to\O^{p_1,\dots,p_A-1,\dots,p_B-1,\dots,p_N}$ given by
\be \label{eta and trace}
\h_{AB}:=\h_{\m\n}\theta_A^\mu\theta_B^\nu\,,\quad tr_{AB}:=\h^{\m\n}\bar\theta_\mu^A\bar\theta_\nu^B\,.
\ee
Note that $\h_{\m\n}$ are the components of the Minkowski metric, while $\h_{AB}$ is the map. According to expression \eqref{local expansion}, this map can also be seen as a bipartite tensor of degree $(1,1)$. In fact, it is an irreducible $(1,1)$ bipartite tensor since its components are represented by the symmetric $2$-tensor $\h_{\m\n}=\h_{(\m\n)}$. 

Reducible and irreducible multipartite tensors can be distinguished using co-trace operators $\s_{AB}:\O^{p_1,\dots,p_A,\dots,p_B,\dots,p_N}\to\O^{p_1,\dots,p_A+1,\dots,p_B-1,\dots,p_N}$ defined by
\be \label{sigma}
\s_{AB}:=-\theta_A^\m \bar\theta^B_\m\,.
\ee
Then, an $N$-partite tensor with $p_{A}\geq p_{B}$ for all $A\leq B$, is said to be in the $GL(d)$-irreducible subspace $\O^{[p_1,\dots,p_N]}\subseteq\O^{p_1,\dots,p_N}$ and is represented by $[\omega]\equiv \o_{[p_1,\dots,p_N]}\in\O^{[p_1,\dots,p_N]}$ if and only if it satisfies the algebraic constraints
\be
\s_{AB}\,\o=0\,,\quad \forall\,A\leq B\qquad \&\qquad
\o^{\top_{AB}}=\o\,,\quad \forall\,A,B\,\,\,\,\,\,\,\,with\,\,\,\,\,\,\,p_A=p_B\,.
\ee
Here, ${}^{\top_{AB}}:\O^{p_1,\dots,p_A,\dots,p_B,\dots,p_N}\to\O^{p_1,\dots,p_B,\dots,p_A,\dots,p_N}$ are transposition maps, acting on any $N$-partite tensor \eqref{local expansion} as a swap of the variables $\theta_A$ and $\theta_B$. 
These constraints constitute the so-called Young symmetry and the $GL(d)$-irreducible subspace $\O^{[p_1,\dots,p_N]}$ is therefore the space of $GL(d)$ Young tableau representations with $N$ columns of respective lengths $p_1$, $p_2$, $\dots$ , $p_N$. Given an arbitrary, reducible $N$-partite tensor $\o\in\O^{p_1,\dots,p_N}$, there is a way to obtain a unique irreducible $N$-partite tensor $[\o]\in\O^{[p_1,\dots,p_N]}$. This is achieved by acting with the Young projector $\mathcal{P}_{[p_1,\dots,p_N]}:\O^{p_1,\dots,p_N}\to \O^{[p_1,\dots,p_N]}$ defined by
\bea\label{2.24}
[\o]:=\mathcal{P}_{[p_1,\dots,p_N]}\,\o~.
\eea 
The Young projector is a polynomial in the $\s$-maps, with fully determined coefficients \cite{BB1,dMH1,BB2,dMH2}.  

Finally, the $\binom{2N}{2}$ algebraic maps \eqref{number}--\eqref{sigma} satisfy certain (anti)commutation relations. These naturally follow from the defining properties \eqref{anticommuting}--\eqref{CCM} and read as
$$[\s_{AB},\s_{BA}]=\hat p_A-\hat p_B\quad,\quad \{\s_{AB},\s_{BC}\}=-\s_{AC}\quad,\quad\{\s_{AB},\s_{CB}\}=0$$
$$[tr_{AB},\h_{AB}]=d-\hat p_A-\hat p_B\quad,\quad\{tr_{AB},\h_{BC}\}=-\s_{CA}\quad,\quad\{\h_{AB},\h_{BC}\}=0\quad,\quad\{tr_{AB},tr_{BC}\}=0$$
\be \label{algebraic maps relations}
[\hat p_A,tr_{AB}]=-tr_{AB}\quad,\quad [\hat p_A,\h_{AB}]=\h_{AB}\quad,\quad [\hat p_A,\s_{AB}]=-[\hat p_B,\s_{AB}]=\s_{AB}
\ee
$$[\s_{AB},\h_{AB}]=0\quad,\quad\{\s_{AB},\h_{AC}\}=0\quad,\quad\{\s_{AB},\h_{BC}\}=-\h_{AC}$$
$$[\s_{AB},tr_{AB}]=0\quad,\quad\{\s_{AB},tr_{AC}\}=-tr_{BC}\quad,\quad\{\s_{AB},tr_{BC}\}=0\,.$$
Maps with all of their indices distinct commute. It is interesting to note that the above maps are related to the $\binom{2N}{2}$ generators of the Lie algebra $\mathfrak{so}(2N)$. 

In addition to the algebraic maps, one can also define differential operators. Exterior derivatives $\dd_A:\O^{p_1,\dots,p_A,\dots,p_N}\to\O^{p_1,\dots,p_A+1,\dots,p_N}$, their co-differentials $\dd^\dagger_A:\O^{p_1,\dots,p_A,\dots,p_N}\to\O^{p_1,\dots,p_A-1,\dots,p_N}$ and the Laplacian are naturally defined as
\be \label{diff operators}
\dd_A:=\theta_A^\m\partial_\m\quad,\quad \dd_A^\dagger:=\bar\theta^A_\m \partial^\m\quad,\quad \Box:=\dd_A\dd^\dagger_A+\dd^\dagger_A\dd_A\,.
\ee
 For distinct labels the above maps commute, i.e. $[\dd_A,\dd^\dagger_{B}]=0$ for $B\ne A$, while both $\dd_A$ and $\dd^\dagger_A$ square to zero for every $A$. Relations between the algebraic and differential maps also follow from \eqref{anticommuting}-\eqref{CCM}. In particular, one finds 
$$\{tr_{AB},\dd_A\}=\dd^\dagger_B\quad,\quad \{tr_{AB},\dd^\dagger_A\}=0\quad,\quad\{\h_{AB},\dd_A\}=0\quad,\quad \{\h_{AB},\dd^\dagger_A\}=\dd_B $$
\be \label{differential maps relations}
\{\s_{AB},\dd_A\}=0\quad,\quad\{\s_{AB},\dd_B\}=-\dd_A\quad,\quad \{\s_{AB},\dd^\dagger_A\}=\dd^\dagger_B\quad,\quad\{\s_{AB},\dd^\dagger_B\}=0\,.\ee
Just like the algebraic operators and their relation to the generators of $\mathfrak{so}(2N)$, the $2N$ differential operators \eqref{diff operators} are related to $2N$ supercharges $Q_A$ defined by \cite{Bonezzi:2018box}
\be \label{supercharges}
Q_A:=\left\{
\begin{array}{ll}
      \frac{-i}{\sqrt{2}}(\dd_A+\dd^\dagger_A) & \quad for\quad A=1,\dots,N \\\\
      \frac{1}{\sqrt{2}}(\dd_A-\dd_A^\dagger) & \quad for\quad A=N+1,\dots,2N\,. \\
\end{array} 
\right.
\ee
These supercharges satisfy $Q^2_A= H$ and $[Q_A,H]=0$, where $H$ is the Hamiltonian $-\frac{\Box}{2}$. Moreover, the supercharges are related to one another by
$$\{Q_A,Q_{N+A}\}=0\qquad\&\qquad [Q_A,Q_{B}]=0\quad \text{for} \quad B\neq N+A\,.$$
Note that typically supercharges anticommute with each other. The apparent difference here is due to the choice of sign convention, which however is more convenient for the considerations of the following section. Had we chosen the alternative convention where the parity $\epsilon(A,B)=1$ in \eqref{anticommuting}, we would indeed find that all supercharges  mutually anticommute. These being said, we can see that \eqref{supercharges} generate a $\mc N=2N$ supersymmetry algebra with $H=-\frac{\Box}{2}$ being the bosonic operator. In addition, the aforementioned $\mathfrak{so}(2N)$ algebra corresponds to its R-symmetry, as can be seen from relations \eqref{differential maps relations}.

\section{Gauge invariant Lagrangians}
Our goal now is to construct Lagrangians for $N$-partite tensor fields in the formalism reviewed in the previous section. As a first step, an inner product $(\cdot,\cdot):\O^{p_1,\dots,p_N}\times\O^{p_1,\dots,p_N} \to\mathbb{R}$ should be defined. The two essential ingredients for this inner product are a Hodge star operator and an integral over degree-1 variables. Starting with the first, we introduce a generalized Hodge star operator $\star_{AB}:\O^{p_1,\dots,p_A,\dots,p_B,\dots,p_N}\to\O^{p_1,\dots,d-p_A,\dots,d-p_B,\dots,p_N}$ defined by
\be\label{Hodge star}
\star_{AB}\o:=\frac{1}{(d-p_A-p_B)!}\,\h_{AB}^{d-p_A-p_B}\o^{\top_{AB}}\,,\quad for \quad d\geq p_A+p_B\,.
\ee
Notice that this operator requires a pair of anticommuting coordinates $\theta_A$ and $\theta_B$ to be defined. On the other hand, the natural integral over anticommuting variables is the Berezin integral
\be 
\int_A\theta_A^{\m_1}\dots\theta_A^{\m_d}\equiv\int d^d\theta_A\,\theta_A^{\m_1}\dots\theta_A^{\m_d}=\epsilon^{\m_1\dots\m_d}\,,
\ee
resulting in the covariant Levi-Civita tensor of Minkowski space. Then, using the floor function, we consider an even number  $k:=2\,\lfloor \frac{N+1}{2}\rfloor$ of anticommuting variables and define the inner product by
\be 
(\o,\omega'):=\int_{1,\dots,k}\o\,\, \star_{12}\star_{34}\cdots\star_{k-1\,\,k}\,\,\omega'\,\equiv\int_{1,\dots,k}\o\,\, \underline{\star}\,\,\omega'\,.
\ee
Obviously, this is well-defined only if $d\geq p_A+p_{A+1}$ for any $A=1,3,5,\dots,k-1$. Moreover, since any scalar function in $\mathbb{R}$ is invariant under the transposition operator ${}^{\top_{AB}}$ for any $A$ and $B$ and the same is true for $\h_{AB}$, the inner product is symmetric, i.e. $(\o,\omega')=(\omega',\o)$. 
\subsection{Kinetic terms}
Consider now a reducible $N$-partite tensor $\o\in\O^{p_1,\dots,p_N}$ and its irreducible counterpart $[\o]=\mc{P}_{[p_1,\dots,p_N]}\o\in\O^{[p_1,\dots,p_N]}$. Irreducible tensors are very important when it comes to physical applications, since it is to such tensors that physical gauge fields correspond. We would like to construct a kinetic term for this tensor, which is manifestly invariant under the infinitesimal  gauge transformation
\be \label{gauge trafo}
\d[\o]=\mc{P}_{[p_1,\dots,p_N]}\sum_{A=1}^N\dd_A\l^{(A)}_{p_1,\dots,p_A-1,\dots,p_N}\,,
\ee
involving reducible gauge parameters $\l^{(A)}$ of degree as indicated in \eqref{gauge trafo}. Note that we restrict here to linear transformations and we are not going to discuss nonlinear gauge theories, such as Yang-Mills theory, in this paper. Moreover, in the spirit of \cite{FS}, we are going to allow the kinetic term to be nonlocal, as long as it leads to local field equations after imposing the gauge fixing condition 
\be \label{gauge fixing}
\dd_A^\dagger[\o]=0\,,\quad \forall\,A\in[1,N]\,.
\ee
Combined with tracelessness, the above condition corresponds to the transverse traceless gauge or to the so-called physical gauge in \cite{BB1,dMH2}. The latter fixes the gauge symmetry completely and, thus, the divergenceless condition \eqref{gauge fixing} is merely a partial gauge fixing.  
The proposed kinetic term that fulfills the aforementioned criteria is 
\be \label{kinetic}
\mc{L}_{(0)}([\o])=\int_{1,\dots,k}\dd_1\dd_3\dots\dd_{k-1} [\o]\,\,\,\underline{\star}\,\,\,\frac{1}{\Box^{\frac{k}{2}-1}}\dd_1\dd_3\dots\dd_{k-1} [\o]\,.
\ee
 This is a generalization of the graded geometric local kinetic term constructed in \cite{CKRS} for the special case of bipartite tensors, as it indeed reduces to it in the case of $k=2$ anticommuting variables, when the inverse box operator disappears. In that case, one obtains a universal kinetic term for bipartite tensors of degree $(p_1,p_2)$, in the sense that the coordinate expression for the corresponding Lagrangian of any field of this type is obtained upon performing the Berezin integration. For example, this includes the scalar kinetic term for $p_1=p_2=0$, the one for differential forms when either $p_1$ or $p_2$ vanishes, the linearised Einstein-Hilbert action prior to gauge fixing when $p_1=p_2=1$ and the Curtright action for the basic hook Young tableau when $p_1=2$ and $p_2=1$. Further details on this point may be found in \cite{CKS}.
 Moreover, it is completely obvious that \eqref{kinetic} is indeed invariant under \eqref{gauge trafo} up to total derivative terms, while it is apparently nonlocal for $N\geq 3$. Despite this nonlocality, the kinetic term is well-defined assuming suitable boundary conditions. Finally, after imposing the condition \eqref{gauge fixing} one obtains local field equations, as desired.

\subsection{Galileon self-interactions}
We now turn our attention to interaction terms. More specifically, we are interested in Galileon self-interactions. These are higher-derivative terms involving just a single gauge field $[\o]$, which have the property that they lead to second order field equations. They were first introduced for scalar fields $\phi$, in which case there is a characteristic Galilean-type symmetry $\d\phi=c+b_{\mu}x^{\mu}$, where $c$ and $b_{\mu}$ are constant. They were subsequently generalised to differential forms \cite{DDE2} and bipartite tensor fields \cite{CKRS}. In all these cases, the corresponding local interaction terms in the Lagrangian are invariant under the same gauge transformation \eqref{gauge trafo} as the kinetic term \eqref{kinetic}. In the following, we briefly review these interactions and provide a new formulation of them as ``generalized kinetic terms'', a notion that will become clear below. In addition, we would like to take this one step further and examine whether, and under which conditions, such interaction terms are possible for bosonic fields of any spin.

\paragraph{A priori gauge invariant interactions.}
Let us first revisit differential forms and bipartite tensors, i.e. when $N=1,2$ and $k=2$, in which cases interactions are local and they were constructed in \cite{CKRS}. The basic interaction term is 
\be \label{bipartite Galileon}
\mc{L}_{(n)}([\o])=\int_{1,2}\dd_1[\o]\left(\dd_1\dd_2[\o]\right)^n\,\,\star_{12}\,\,\dd_1[\o]\left(\dd_1\dd_2[\o]\right)^n.
\ee
Note that the case of scalar field, when $N=0$ and $k=0$, also belongs in this class. In addition, one can see that the above inner product is well-defined for $d\geq (n+1)(p_1+p_{2}+1)+n$. Thus, these Galileon interactions are defined for $n\leq n_{max}=\lfloor\frac{d+1}{p_1+p_2+2}\rfloor-1$. In other words, the general Galileon Lagrangian which should be added to the kinetic term is 
\be 
{\cal L}_{\text{Gal}}([\omega])=\sum_{n=1}^{n_{max}}{\cal L}_{(n)}([\omega])\,.
\ee 
Note also that the very symmetric and suggestive expression \eqref{bipartite Galileon} did not appear as such in \cite{CKS}. In this new form, it becomes clear that all bipartite Galileons are simply given by the inner product of the quantity $\dd_1[\o]\left(\dd_1\dd_2[\o]\right)^n$ with itself, and thus the full kinetic plus interaction Lagrangian simply becomes
\be \label{generalised kinetic}
\mc{L}([\o])=\sum_{n=0}^{n_{max}}\int_{1,2} \dd_1 [\o]_{(n+1)}\star_{12}\dd_1 [\o]_{(n+1)}\,,
\ee 
where we defined $[\o]_{(n+1)}\equiv[\o]\left(\dd_1\dd_2[\o]\right)^n$.
This highlights the importance of the generalised Hodge star operator $\star_{12}$. Most importantly, it shows that the totality of the kinetic plus higher-derivative Galileon interaction terms for any type of bipartite tensor is captured by the ``generalised kinetic term'' \eqref{generalised kinetic}. 

In fact, the totality part of the latter statement is only true for $p_1\ne p_2$. In case the two partial degrees are equal, $p_1=p_2=p$, as happens for instance for scalar fields ($p=0$) or the graviton ($p=1$), there exist additional higher-derivative interactions with the same properties as above. These are given as 
\be\label{bipartiteGal0}
{\widetilde{\cal L}_{\text{Gal}}([\o])=\sum_{n=1}^{n_{\text{max}}}\int_{1,2} \,\eta^{D-k_n}\,[\o]\,(\dd_1\dd_2\, [\o])^{n},}
\ee
with $k_{n}=(p+1)n+p$ and $n_{max}=\lfloor \sfrac {d-p}{p+1}\rfloor$. Unlike the interaction given in \eqref{bipartite Galileon}, these additional ones allow for an odd number of fields to appear. As such, they cannot be written in the form of a generalised kinetic term, but they are otherwise permissible. 

Returning to the question of spins higher than 2, our goal is to find a $k>2$ extension of \eqref{bipartite Galileon}, such that it is compatible with \eqref{kinetic} for $n=0$ and it is invariant under \eqref{gauge trafo}. Once this is achieved, the remaining question is whether it leads to local field equations after imposing \eqref{gauge fixing}. 
The unification of kinetic and interaction terms by means of \eqref{generalised kinetic} already suggests a candidate.  
Indeed, since the bipartite kinetic term is simply the ``degenerate'' interaction term for $n=0$, 
it is reasonable to think that in the multipartite case we should consider the substitution $[\omega] \to [\omega] \left(\frac{1}{\Box^{\frac{k}{2}-1}}\dd_1\dots \dd_k[\omega]\right)^n$. Using the shorthand notation $D:=\dd_1\dots \dd_k$ for the full higher order differential and $D_{\text{odd}}:=\dd_1\dd_3\dots\dd_{k-1}$, we can then define the enhanced expression{\footnote{We do not introduce a separate notation with respect to the bipartite case, since the latter just follows for $k=2$.}} 
\be 
[\omega]_{(n+1)}:= [\omega] \left(\frac{1}{\Box^{\frac{k}{2}-1}}D[\omega]\right)^n\,,
\ee 
and write down the following interaction terms:
\be \label{galileon multi}
{\cal L}_{\text{Gal}}([\omega])=\sum_{n=1}^{n_{max}}\mc{L}_{(n)}([\o])=\sum_{n=1}^{n_{max}}\int_{1,\dots,k} D_{\text{odd}} [\o]_{(n+1)}\,\underline{\star}\,\,\,\frac{1}{\Box^{\frac{k}{2}-1}}D_{\text{odd}} [\o]_{(n+1)}.
\ee
These contain $2n+2$ appearances of the field and are well-defined only for
\begin{equation}
        d\geq (n+1)(p_A+p_{A+1}+1)+n\,,\quad \forall\, A=1,3,5,\dots,k-1\,.
\end{equation}
Thus, $n_{max}$ can be determined by considering the maximum integer $n$ obeying all the above $k/2$ inequalities. 

At this stage, one has to also make sure that the above interactions do not correspond to total derivative terms. This is very easy to be checked in the graded formalism. As was already observed in \cite{CKRS} for the bipartite case, these interactions exhibit a property that could be termed ``evenophilia''. This means that for bipartite tensor fields with odd total degree $P$ they correspond to total derivatives. It is simple to check that this property continues to hold in the multipartite case, since the quantity $D[\o]$ appearing in \eqref{galileon multi} for $n\geq 1$ has odd grading for $P$ odd. Then, performing some partial integrations in \eqref{galileon multi} leads to the appearance of $(D[\o])^{n+1}$, which obviously vanishes for $n\geq 1$. Thus genuine interactions that do not correspond to total derivatives are the ones when the total degree $P$ of the field is even. 

As in the bipartite case, in addition to the above candidates for Galileon interactions, an enhancement takes place when all partial degrees are equal, namely when $p_{A}=p$ for all $A$. Then the potential gauge invariant Galileon interactions contain $n+1$ appearances of the field, which is not necessarily an even number as before. These have the form
\be
\label{Galhigherspin0}
\widetilde{\mc{L}}_{(n)}([\o])=\int_{1,\dots,k}\, H^{\{d-(n-1)(1-p),\,np\}}\,[\o]\,\left(\frac{1}{\Box^{\frac{k}{2}-1}}D[\o]\right)^{n},
\ee
where we have defined the quantity $H^{\{d,p\}}
:=\prod_{A=1,3,5,...}^{k-1}\eta^{d-p_A-p_{A+1}-1}_{A\,A+1}$. For $n=1$ they reduce to the respective kinetic term \eqref{kinetic}, while for higher odd values of $n$ they can be brought to the form \eqref{galileon multi} after some partial integrations. For even $n$, the above interaction terms cannot be identified with any term in \eqref{galileon multi}, as happened in the bipartite case.

Finally, one can easily see that both \eqref{galileon multi} and \eqref{Galhigherspin0} are a priori gauge invariant under \eqref{gauge trafo}. The only thing that remains to be checked is whether they lead to local field equations upon imposing \eqref{gauge fixing}. The answer to this last question is negative and we will illustrate this in the simple example of a spin-$4$ field in five dimensions. This is described by the irreducible $4$-partite tensor $\o_{[1,1,1,1]}$, with its components being a fully symmetric $4$-tensor. The first candidate Galileon interaction term $\widetilde{\mc{L}}_{(2)}$ has three field appearances and it reads as 
\be \label{special case}
\widetilde{\mc{L}}_{(2)}([\o])=\int_{1,2,3,4}[\omega]\left(\frac 1{\Box}\,\dd_1\dd_2\dd_3\dd_4[\o]\right)^2~.
\ee  
Expanding in components, one obtains directly 
\begin{equation*}
\widetilde{\mc{L}}_{(2)}([\o])=\epsilon^{\m[5]}\epsilon^{\n[5]}\epsilon^{\k[5]}\epsilon^{\l[5]}
\,\o_{\m_1\n_1\k_1\l_1}\,\frac 1{\Box}\,\partial_{\m_2}\partial_{\n_2}\partial_{\k_2}\partial_{\l_2}\o_{\m_3\n_3\k_3\l_3}\, \frac 1{\Box}\,\partial_{\m_4}\partial_{\n_4}\partial_{\k_4}\partial_{\l_4}\o_{\m_5\n_5\k_5\l_5}~,
\end{equation*}  
where $\e^{\m[5]}=\epsilon^{\m_1\m_2\m_3\m_4\m_5}$, etc. The condition \eqref{gauge fixing} can also be written in local form $\partial^{\m}\omega_{\m\n\k\l}=0$.
After imposing it, one can see that several terms in the Lagrangian are instantly dropped. In several other terms the inverse boxes will cancel as desired, but there will still be terms where the inverse boxes do not cancel. These will be of the general form 
\be \label{example}
\omega\, \frac {1}{\Box}\partial_{\m}\partial_{\n}\partial_{\k}\partial_{\l}\,\o\,\frac 1{\Box}\partial^\m\partial^\n\partial^\k\partial^\l\,\o~,
\ee   
namely ones with multiple cross-contracted derivatives. Note also that this issue cannot be ameliorated by partial integration. One could imagine that applying the full transverse traceless (or physical) gauge condition would maybe lead to locality. This is also not the case, since there will still exist terms \eqref{example} that do not contain any traces of $\o$. 

Therefore we conclude that, even though gauge invariant, the candidate Lagrangians \eqref{galileon multi} and \eqref{special case} do not lead to local field equations after the partial gauge fixing \eqref{gauge fixing}; not even full gauge fixing is adequate. Thus, requiring invariance under \eqref{gauge trafo} leads to a nonlocal Lagrangian for $N\geq 3$. Allowing for nonlocality would indicate that one may just ignore the inverse boxes in the above Lagrangians. However, then the resulting field equations would be of order higher than two in derivatives and ghost instabilities are expected to arise. 

\paragraph{A priori local interactions.} Following a different strategy, one could attempt to avoid the above problem of nonlocality by considering Lagrangians that do not contain inverse box operators to start with. Rather than discussing this in general, let us look at a specific illustrative example. Consider an irreducible $4$-partite tensor $\o_{[3,2,2,1]}$ in seven dimensions. 
Then, one can write an interaction term of the type 
\be \label{hsint2}
\int_{1,2,3,4} [\o]\,\dd_3\dd_4[\o]^{\top_{12}}\dd_3\dd_4\left(\left([\o]^{\top_{34}}\right)^{\top_{23}}\right)^{\top_{13}}.
\ee 
 At face value, this term does not vanish because the sum of field degrees is even (eight) and the four different odd variables are saturated in the integration due to the different transpositions that are available for $N\geq 3$. Moreover, the term leads obviously to field equations with only two derivatives acting on the field, without the need for any additional ingredients. 
 
 However, although the existence of this type of terms is interesting in itself, not all transformations \eqref{gauge trafo} are  symmetries of this Lagrangian any longer. It is clear that this will be the case for any type of interaction constructed in the spirit of \eqref{hsint2}, since there are less derivatives than required for the symmetries to be preserved.
We have thus given strong supporting evidence  that higher derivative Galileon self-interactions leading to local, second order field equations and respecting the gauge symmetries \eqref{gauge trafo} cannot be constructed this way for $N\ge 3$.

\section{Additional results and conclusions}\label{sec4}
Apart from the setting considered in this paper, it is worth mentioning that the formalism reviewed here can also be very useful in studying other aspects of gauge theories, such as dualities. For example, one can use it to construct first-order Lagrangians in order to study off-shell Electric/Magnetic duality between tensor fields. In the spirit advocated here, one expects to find a universal first-order Lagrangian that captures all kinds of dualities between certain tensor fields. Indeed, such a Lagrangian was constructed in \cite{CKS} for certain $N=2$ (bipartite) tensors and it reads as
\be\label{master}
{  \mathcal{L}^{(p_1,p_2)}(F,\l)=\int_{1,2}F_{p_1,p_2}\star_{12} \mathcal{O} \,F_{p_1,p_2}+\int_{1,2}\dd_1 F_{p_1,p_2}\ast_1\ast_2\,\lambda_{p_1+1,p_2}}\,,
\ee
for $d\geq p_1+p_2+1$. Here, $F$ and $\l$ are independent reducible bipartite tensors and the star operators $\ast$ correspond to the standard Hodge operators acting on differential forms. Standard here means that $\ast_1$ acts on the first column of $\l$ and $\ast_2$ acts on the second, unlike the generalised operator $\star_{12}$ that acts on both and it is distinct from their combined action (see \cite{CKS} for a detailed discussion).
Moreover, ${\cal O}$ is an operator acting on bipartite tensors such that $\mc O \,\dd_1\o_{p_1-1,p_2}:=\dd_1\o_{[p_1-1,p_2]}+\dd_2(\dots)$.
This operator can be fully determined by the above defining property and it yields the kinetic term for an irreducible potential $\o_{[p_1-1,p_2]}$ upon integrating out $\l$ using its equation of motion. On the other hand, integrating out $F$ leads to another gauge theory for a field $\widetilde{\omega}$ dual to $\omega_{[p_1-1,p_2]}$.  

The above first-order Lagrangian encapsulates all possible dual theories for differential forms and generalized gravitons, which are irreducible bipartite tensor fields of degree $(p,1)$. The dual field one obtains depends on the parameters $p_1$ and $p_2$. Specifically, for $p_2=0$, the original and dual fields are  $\omega_{p_1-1}$ and $\widetilde{\omega}_{d-p_1-1}$ respectively, both being differential forms. This is standard electric/magnetic duality. For $p_2=1$, the corresponding fields are $\omega_{[p_1-1,1]}$ and $\widetilde{\omega}_{[d-p_1-1,1]}$, corresponding to standard duality between two generalised gravitons. In case $p_1=1$, the fields are $\omega_{p_2}$ and $\widetilde{\omega}_{[d-2,p_2]}$, a duality ofter called exotic, since it related a differential form to a bipartite tensor. Finally, for $p_1=2$ one obtains $\omega_{[1,p_2]}$ and $\widetilde{\omega}_{[d-3,p_2]}$.  
It would  be interesting to extend this result and study the off-shell E/M dualities involving multipartite tensors for arbitrary $N$. 

In addition, one can use this formalism to study on-shell dualities and compare the so-called standard and exotic dual fields. A first step towards this direction was made in \cite{CK}, where all possible duals of differential forms were analysed on-shell and it was found that only a fraction of them are algebraically independent. This analysis is based on the preceding one 
for gravitons, which was performed in \cite{Henneaux:2019zod} and led to the result that the double dual of the graviton is not independent to the original field.

In a different direction, it is worth mentioning that a number of generalizations of single field, flat spacetime Galileons also exist (see \cite{Deffayet:2013lga} for a review). These generalizations are also elegantly described in the formalism of graded geometry, with the necessary minimal modifications \cite{CKRS}. More specifically, it is simple to extend \eqref{bipartite Galileon} to a Lagrangian for arbitrary number of species of bipartite tensor fields. An additional property in that case is that also fields with odd total degree may participate in cross-species interactions, even though they cannot self-interact.  Furthermore, it is simple to contruct Lagrangians whose field equations are up to second order in derivatives instead of strictly second order. Finally, curved spacetime generalizations of Galileons exist too. Although for scalars and differential forms this is relatively straightforward by addition of suitable counter terms, the situation is more tricky for bipartite tensors. Indeed in the simplest case one would be dealing with two species of gravitons, which however cannot have nontrivial cross interactions \cite{Boulanger:2000rq}. This issue was discussed in \cite{CKRS}, where a proposal for five-dimensional bigravity based on Gauss-Bonnet terms was put forward. It would be interesting to revisit this approach in future work.    

In conclusion, we have discussed a number of applications of graded geometry in field theory for different types of fields. Even in the simple setting described here, this formalism offers an elegant geometric way to unify different Lagrangians and underline their common characteristics.  
 This set up is particularly effective when studying mixed-symmetry tensors, due to an isomorphism relating smooth functions on the graded supermanifold $\mc M=\oplus^NT[1]M$ with such tensors in $M$. A generalised Hodge star operator is naturally defined and it leads to an inner product that can be used to write down kinetic terms, as well as higher-derivative  self-interactions for arbitrary mixed-symmetry tensors. Such non-trivial, gauge invariant, local interactions are possible for all tensors with $N\le 2$ and even total degree. On the other hand, for $N\ge 3$, we provided strong evidence that such potential self-interactions cannot be fully gauge invariant and local at the same time.

\acknowledgments{
We would like to thank the European Institute for Sciences and
their Applications (EISA) in Corfu and in particular Ifigeneia Moraiti and George Zoupanos. 
The work of A.Ch. and G.K. is supported by the Croatian Science
Foundation Project ``New Geometries for Gravity and Spacetime'' (IP-2018-01-7615) and also partially supported by the European Union through the European Regional Development Fund - The Competitiveness and Cohesion Operational Programme (KK.01.1.1.06).}

\end{document}